\documentclass[]{spie}  %>>> use for US letter paper
\usepackage{amsmath,amsfonts,amssymb}
\usepackage{graphicx}
\usepackage[colorlinks=true, allcolors=blue]{hyperref}
\usepackage{multirow}
\usepackage{subfig}
\usepackage{graphicx}
\usepackage{microtype}
\usepackage{array}
\newcolumntype{P}[1]{>{\centering\arraybackslash}p{#1}}
\newcolumntype{M}[1]{>{\centering\arraybackslash}m{#1}}

\title{Multi-layer anti-reflection coats using ePTFE membrane for mm-wavelength plastic optics}

\author[a]{Miranda~Eiben}
\author[a]{Keara~Carter}
\author[a]{Marion~Dierickx}
\author[a]{Brodi~Elwood}
\author[a]{Paul~Grimes}
\author[a]{John~Kovac}
\author[b]{Matthew~Miller}
\author[a]{Matthew~A.~Petroff}
\author[a]{Annie~Polish}
\author[a]{Clara~Vergès}

\affil[a]{Center for Astrophysics $\vert$ Harvard \& Smithsonian, 60 Garden St, Cambridge, MA 02138, USA}
\affil[b]{Department of Physics and Astronomy, Brigham Young University, Provo, UT 84602, USA}

\authorinfo{Further author information: (Send correspondence to Miranda Eiben)\\ E-mail: miranda.eiben@cfa.harvard.edu}
\begin{document} 
\maketitle

\begin{abstract}
Future millimeter wavelength experiments aim to both increase aperture diameters and broaden bandwidths to increase the sensitivity of the receivers. These changes produce a challenging anti-reflection (AR) design problem for refracting and transmissive optics. The higher frequency plastic optics require consistently thin polymer coats across a wide area, while wider bandwidths require multilayer designs. We present multilayer AR coats for plastic optics of the high frequency BICEP Array receiver (200-300 GHz) utilizing an expanded polytetrafluoroethylene (ePTFE) membrane, layered and compressively heat-bonded to itself. This process allows for a range of densities (from 0.3g/cc to 1g/cc) and thicknesses ($>$0.05mm) over a wide radius (33cm), opening the parameter space of potential AR coats in interesting directions. The layered ePTFE membrane has been combined with other polymer layers to produce band average reflections between 0.2\% and 0.6\% on high density polyethylene and a thin high modulus polyethylene window, respectively.
\end{abstract}


\section{INTRODUCTION}
\label{sec:intro}  % \label{} allows reference to this section

% \begin{enumerate}
%     \item Why do we want to observe in the millimeter?
%     \item Why do we need AR coats?
%     \item Why are AR coats hard?
% \end{enumerate}
The millimeter wavelength regime (from 300 to 30 GHz, or 1 mm to 1 cm wavelengths) provides a rich source of a variety of astrophysical phenomena. Astrophysical sources range from the very close, such as objects within the solar system, to the oldest light we can see, the Cosmic Microwave Background (CMB). The atmospheric absorption lines within the millimeter wavelengths, however, limit commercial viability and therefore the products and materials designed for use in mm-wavelength optics.  

Anti-reflection (AR) coats are a critical component of an optical system, reducing light lost to reflections and preventing spurious signal from reaching detectors. AR coats utilizing layered materials are typically quarter wavelength electrical lengths to produce destructive interference. Millimeter wavelength AR coats therefore exist in a unique macroscopic-but-still-thin regime; quarter wavelength optical lengths range from 0.3 mm to 3 mm for 300 to 30 GHz. 

Instrumentalists designing AR coats for mm-wavelength optics must therefore source or produce materials that have the appropriate optical properties over the widths of their transmissive optics. Astronomical mm receivers are typically cryogenic, requiring (at least) vacuum windows and IR blocking filters \cite{Grimes2020a,SeanBryan2018,Hui2018BA,bicep3,abazajian2019cmbs4,Thornton2016ActPol,Rhoades2018}. The CMB community also produces many refracting receivers with lenses, calling for AR coating solutions for lenses as well. \cite{Hui2018BA,bicep3,abazajian2019cmbs4,Thornton2016ActPol,Rhoades2018}. Materials used for these transmissive optics are typically various polymers, such as many forms of polyethylene (PE), polypropylene (PP), polytetraflouroethylene (PTFE, commercial name Teflon), and nylon \cite{abazajian2019cmbs4,Shitvov2022}. Other common materials are alumina ceramic, which has been used in the past for lenses and filters, and silicon, also used for lenses \cite{bicep3,Dierickx2021,Rhoades2018,Thornton2016ActPol}. Each material has unique challenges associated with producing and generating AR coats; alumina, for instance, has a high index of refraction (approximately 3.1) and a much smaller coefficient of thermal expansion (CTE) than most materials used to adhere AR coats in addition to being notoriously difficult to machine \cite{Golec_2020,Jeong_2023}. Plastic optics may have a better match to the CTE of the polymers used for hot melt adhesives, but due to their relatively low index (generally between 1.4 to 1.8) require even lower index materials to coat them, which may be difficult to source or produce \cite{Lamb1996,Dierickx2021,eiben2022laminate,Shitvov2022}.

Other ways to eliminate reflection are possible, such as machined pyramids to smoothly transition light into the optic, or depositing intermediate index material directly onto the optic to smoothly increase the impedance into the optic \cite{Golec_2020,Jeong_2023}. Active research is being conducted in these areas, with exciting possibilities for future optics. For this proceedings, the authors will focus on the parameter space provided by layered AR coats, as there remain optics where these newer solutions may require significant research and development effort (such as the vacuum window).

The layered AR coats for two mm/sub-mm wavelength experiments will be covered in this proceedings. BICEP Array (BA) is a set of four CMB refracting receivers at the South Pole. The receivers have a large open aperture: the PE vacuum windows require AR coats at least 70 cm wide to cover the illuminated area, and the filters and lenses require similar. Previous similarly scaled receivers deployed are BICEP3 (95 GHz), BA1 (35 GHz), and BA2 (150 GHz) \cite{Hui2018BA,bicep3}. The fourth BA receiver (BA4) is the highest frequency receiver (200--300 GHz) and is anticipated to deploy this austral summer to South Pole \cite{nakato2024}. Additionally, a wide-band upgrade to the Sub-Millimeter Array (wSMA) is actively being developed. The SMA is a interferometer on Mauna Kea with eight antennas. The wSMA upgrade is redesigning the cryogenic receivers within each antenna and therefore the vacuum window for each receiver is being redesigned. wSMA has a very large high frequency bandwidth from 190 to 380 GHz, but has a much smaller scale than BA receivers. The illuminated area of the window is only 10.7 cm wide \cite{Grimes2020a}.

In this proceedings, we will discuss the theoretical ideal layered AR coats over a bandwidth and how achieved AR coats compare to the ideal. Then we will discuss how high frequency AR coats are particularly challenging, a useful expanded PTFE (ePTFE) membrane for high frequency AR coats, and the design process for BA4 AR coats and the wSMA window. We will conclude with the potential future of these technologies.

\section{IDEAL AR COATS}
% The goal of anti-reflection is to reduce the reflected power from the impedance change into an optic over the observed bandwidth. The best theoretical way to eliminate reflection is to provide an even ramp of impedance change into the core material. Without a sharp jump in impedance, the wave will gradually transition into the core material. If one is trying to ease the transition with a finite number of layers, however, one must take a different approach. With a single layer it is possible to produce perfect destructive interference at a single wavelength, with two layers perfect destructive interference at two wavelengths, etc. This can be seen in Figure \ref{fig:cheb_ref}. The placement of those `nulls' of destructive interference in frequency space sets the reflection profile over the band.

\begin{figure}
    \centering
    \includegraphics{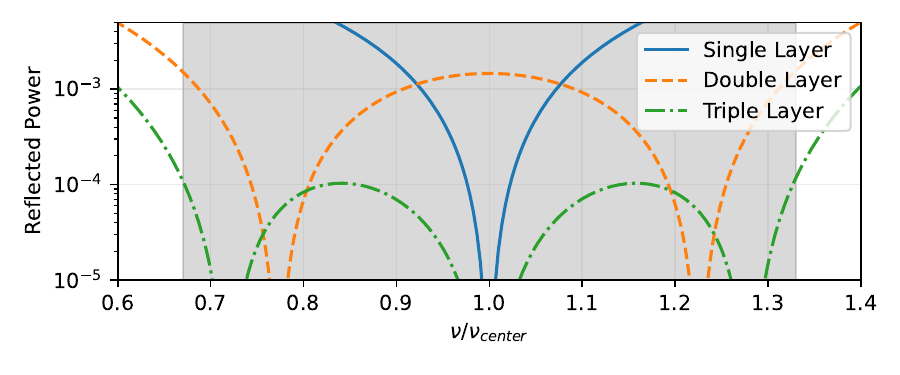}
    \caption{Reflections from ideal single, double, and triple layer anti-reflection coats for HDPE (n=1.55) over a fractional bandwidth of 0.66 (as required for wSMA) \cite{Grimes2020a}}
    \label{fig:cheb_ref}
\end{figure}

The goal of anti-reflection is to reduce the reflected power from the impedance change into an optic over the observed bandwidth. Ideal AR performance can be approached with a very large number of quarter wavelength layers with small changes in index between adjacent layers. However, for practicality we require a minimal number of layers, and careful optimization of the number of layers and desired performance is required. With a single layer it is possible to produce perfect destructive interference at a single wavelength, with two layers perfect destructive interference at two wavelengths, etc. This can be seen in Figure \ref{fig:cheb_ref}. The placement of those `nulls' of destructive interference in frequency space sets the reflection profile over the band.

There are two design directives that one may take when exploring the ideal AR coat, related to the design goal. One may try to keep the reflections as flat as possible, or as low as possible on average over the band. We choose to try to keep the band average reflection as low as possible over a top-hat band as the ideal case for our optics. Therefore, the ideal finite layered AR coat should move nulls to keep the maximum reflections even within the band. An identical problem is found in designing quarter wave transformers, as described in \textit{Microwave Engineering} by Pozar \cite{PozarDavidM1997Me}. We can find the idealized solution by using Chebyshev polynomials to identify the appropriate indexes for a given bandwidth, using the exact same approach that Pozar does with transformers. 

\subsection{Chebyshev Polynomials}
Chebyshev polynomials are `equal ripple' polynomials over a bandwidth. We can find the parameters of a multi-layer quarter wavelength AR coat by solving for the partial reflections off each layer. Chebyshev polynomials follow the formula:

\begin{align}
    T_n (x) = 2xT_{n-1}(x) - T_{n-2}(x),
\end{align}
where $n$ is the order of the polynomial. The first three go like:

\begin{align}
    T_1(x) &= x \\
    T_2(x) &= 2x^2 - 1 \\
    T_3(x) &= 4x^3 - 3x.
\end{align}

Chapter 5.7 of \textit{Microwave Engineering} describes how one may utilize Chebyshev polynomials to produce even ripple reflections across a bandwidth \cite{PozarDavidM1997Me}. We will summarize the final design steps of an even and odd number of AR layers in the next section.

\subsubsection{Example Chebyshev solution for double and triple layer AR coats}

Starting with the formula laid out in Chapter 5.7 of \textit{Microwave Engineering}, each solution requires initially calculating the bandwidth associated with the maximum ripple height \cite{PozarDavidM1997Me}. To calculate this, we iteratively test the bandwidth equation with an array of potential maximum ripple heights:

\begin{align}
      \sec{\theta_m} &\approx \cosh\left[\frac{1}{N}\cosh^{-1}\left(\left|\frac{\ln{n_c}}{2A}\right|\right)\right],
\end{align}
where $\theta_m$ is the bandwidth in phase space, $N$ is the number of layers, $n_c$ is the core index, and $A$ is the maximum reflected power ripple height.

For a double layer AR coat, we start with the polynomial:

\begin{align}
    2\Gamma_0 \cos{2\theta} + \Gamma_1 &= A\sec^2{\theta_m}+A\sec^2{\theta_m}\cos{2\theta}-A. 
\end{align}

If we equate similar terms, we can solve for each of the partial reflections:

\begin{align}
      2\Gamma_0 &= A\sec^2{\theta_m}\\
      \Gamma_1 &= A\sec^2{\theta_m}-A,
\end{align}

which results in ideal indexes for each of the layers:

\begin{align}
    n_1 &= Z_0 \exp{\left[-\ln{\frac{Z_0}{n_2}}+2\Gamma_1\right]} \\
    n_2 &= Z_0 \exp{\left[-\ln{\frac{Z_0}{n_c}}+2\Gamma_0\right]},
\end{align}
where $Z_0$ is the vacuum impedance.

Essentially the exact same procedure is followed for triple layer AR coats, with the initial polynomial starting as:

\begin{align}
    2\Gamma_0 \cos{3\theta} + 2\Gamma_1 \cos(\theta)&=   A\sec^3{\theta_m}\cos(3\theta)+3A\sec^3{\theta_m}\cos{\theta}-3A\sec{\theta_m}\cos{\theta}.
\end{align}

The partial reflections become:

\begin{align}
      2\Gamma_0 &= A\sec^3{\theta_m}\\
      2\Gamma_1 &= 3A(\sec^3{\theta_m}-\sec{\theta_m}).
\end{align}

Because of symmetry, we also assume that $\Gamma_3 = \Gamma_0$ and $\Gamma_2 = \Gamma_1$. Finally the indexes are:

\begin{align}
    n_1 &= Z_0 \exp{\left[-\ln{\frac{Z_0}{n_2}}+2\Gamma_2\right]} \\
    n_2 &= Z_0 \exp{\left[-\ln{\frac{Z_0}{n_3}}+2\Gamma_1\right]} \\
    n_3 &= Z_0 \exp{\left[-\ln{\frac{Z_0}{n_c}}+2\Gamma_0\right]}.
\end{align}

\begin{table}[]
    \centering
    \begin{tabular}{|M{1.8cm}|M{4.2cm}|M{4.3cm}|M{4.4cm}|}
         \hline
        \rule[-1ex]{0pt}{3.5ex} \textbf{Layer} & \textbf{BICEP3} & \textbf{BA4} & \textbf{wSMA} \\
        \rule[-1ex]{0pt}{3.5ex}  & ($\nu_{cen}$=95 GHz, $\frac{\Delta \nu}{\nu_{cen}}$ =0.25) & ($\nu_{cen}$=250 GHz, $\frac{\Delta \nu}{\nu_{cen}}$ = 0.4) & ($\nu_{cen}$=285 GHz, $\frac{\Delta \nu}{\nu_{cen}}$ = 0.66) \\
         \hline
        \rule[-1ex]{0pt}{3.5ex} \textbf{Single} & $n_1=1.245$, $t_1=634 \mu$m & $n_1=1.245$, $t_1=241\mu$m  & $n_1=1.245$, $t_1=211\mu$m  \\
         \hline
        \rule[-1ex]{0pt}{3.5ex} \multirow{2}{*}{\textbf{Double}} & $n_1=1.118$, $t_1=706\mu$m & $n_1=1.122$, $t_1=267\mu$m & $n_1=1.133$, $t_1=232\mu$m \\
        \rule[-1ex]{0pt}{3.5ex}  & $n_2=1.386$, $t_2=569\mu$m & $n_2=1.382$, $t_2=217\mu$m & $n_2=1.368$, $t_2=192\mu$m \\
            \hline
        \rule[-1ex]{0pt}{3.5ex} \multirow{3}{*}{\textbf{Triple}} & $n_1=1.058$, $t_1=746\mu$m & $n_1=1.061$, $t_1=283\mu$m & $n_1=1.070$, $t_1=246\mu$m \\
        \rule[-1ex]{0pt}{3.5ex}  & $n_2=1.245$, $t_2=634\mu$m & $n_2=1.245$, $t_2=241\mu$m & $n_2=1.245$, $t_2=211\mu$m \\
        \rule[-1ex]{0pt}{3.5ex}  & $n_3=1.465$, $t_3=539\mu$m & $n_3=1.461$, $t_3=205\mu$m & $n_3=1.449$, $t_3=182\mu$m \\
         \hline
    \end{tabular}
    \caption{Chebyshev single, double, and triple anti-reflection coat layer parameters on HDPE (n=1.55) for BICEP3 ($\nu_{center}$=95 GHz, fractional bandwidth=0.25), BA4 ($\nu_{center}$=250 GHz, fractional bandwidth=0.4), and wSMA ($\nu_{center}$=285 GHz, fractional bandwidth=0.66). }
    \label{tab:cheb_param}
\end{table}

In Table~\ref{tab:cheb_param}, we report the found Chebyshev indexes and thicknesses for single, double, and triple layer AR coats of HDPE for BICEP3 ($\nu_{center}$=95 GHz, fractional bandwidth=0.25), BA4 ($\nu_{center}$=250 GHz, fractional bandwidth=0.4), and wSMA ($\nu_{center}$=285 GHz, fractional bandwidth=0.66). Note that the Chebyshev polynomials only change the index of a layer; the thickness changes to keep the layer at a quarter wavelength electrical length at the center frequency. For wider bandwidths, the multi-layer AR coats move the indexes slightly toward the $\sqrt{n_{core}}$ (the upper layer indexes go up, the deeper layer indexes go down). These impedance changes move the nulls away from the center frequency, therefore allowing a higher ripple.

\subsection{Ideal AR Coats Compared to Achieved AR Coats}

Practically, it is difficult to find materials that have precisely the properties required by the Chebyshev polynomials: for instance, in Table~\ref{tab:cheb_param}, the double layer Chebyshev solution for BA4 calls for a second layer with properties $n_2=1.382$, $t_2=217 \mu$m. The index is possible to generate (potentially with a relatively high density sintered PTFE), but the thickness is difficult to acquire. Additional necessary adhesive layers will also change the effective electrical length at high frequencies. A typical adhesive layer such as 2 mil (50.8 $\mu$m) LDPE has an electrical length of 76.7 $\mu$m: equivalent to approximately 1/8 $\lambda$ at 250 GHz ($\lambda$=1200 $\mu$m), which is not insignificant! Adhesion layers should be included when possible in the design process, as they may additionally help anti-reflection or may change the best solutions for other layers.

We compare the achieved AR coats on recent optics on the current generation of BICEP/\emph{Keck} instruments (and a few other instruments) to the ideal Chebyshev AR coats. Recent process improvements for the ePTFE for plastic optics have allowed for better control of the final AR layers produced, such that the single layer AR coats for BA2 produce reflections close to the ideal limit \cite{Dierickx2021,eiben2022laminate}. Some of the improved performance at wider bandwidths are related to new materials and new design process. The high frequency (BA4 and wSMA) AR coats will be discussed in Section~\ref{sec:high_freq}.

\begin{figure}
    \centering
    \includegraphics[width=15cm]{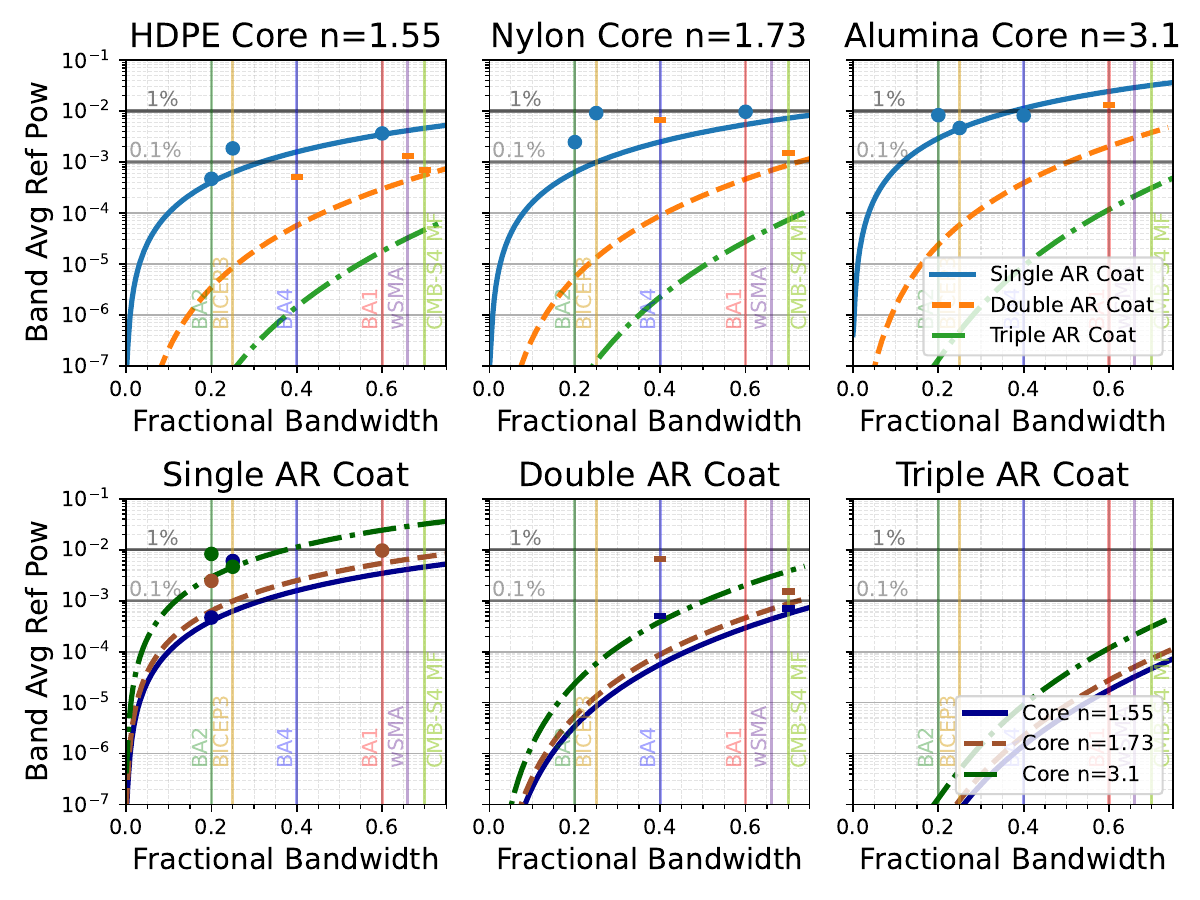}
    \caption{Average in-band reflections for ideal anti-reflection coats on infinitely thick cores of HDPE (n=1.55), nylon (n=1.73), and alumina (n=3.1). Points represent achieved AR coats for each material on each instrument where shape denotes rough equivalent number of layers. Circles are single layer equivalent AR coats, lines double layers.}
    \label{fig:cheb_comp}
\end{figure}

The three materials (HDPE, nylon, and alumina) are the three materials within the BA receivers that require layered AR coats. BA lenses and windows are made of PE, the 50 K filter is made of alumina, and the 4 K filter is made of Nylon 6 \cite{Dierickx2021}. One instrument (wSMA) is not a wide aperture CMB receiver but rather a wide bandwidth interferometer receiver \cite{Grimes2020a}; we are collaborating with the wSMA team to make a high modulus polyethylene (HMPE) AR coated window for their new wider bandwidth. An HMPE window production update was published in the last SPIE, and a full report on HMPE windows is in preparation \cite{eiben2022laminate}. The CMB-S4 Mid-Frequency (MF) optics (combined 95 GHz and 150 GHz bands) are expected to be tested within the PreSAT receiver \cite{petroff2024}. The CMB-S4 MF AR coats in Figure~\ref{fig:cheb_comp} are proposed coats with known materials.

\begin{table}
\caption{Generated AR coats for each recent instrument in the BICEP/Keck collaboration, and two external collaborators (wSMA and CMB-S4 MF). BICEP Array (BA) bands should be similar to CMB-S4 bands (BA1 to CMB-S4 LF, BA4 to CMB-S4 HF). All AR layers' electrical lengths are summed and reported for each material, and those are used to estimate the equivalent number of quarter wavelength layers.} 
\label{tab:AR_params}
\begin{center}       
\begin{tabular}{|M{2.5cm}|M{2cm}|M{2cm}|M{2.1cm}|M{2.3cm}|M{2cm}|M{2cm}|} 
\hline
\rule[-1ex]{0pt}{3.5ex} \textbf{Instrument} & \textbf{Center Freq [GHz]} & \textbf{Material} & \textbf{Quarter Wavelength [mm]} & \textbf{Summed Actual AR Electrical Length [mm]} & \textbf{Estimated Equivalent Layers}  \\
\hline
\rule[-1ex]{0pt}{3.5ex}  \multirow{3}{*}{BA2} & \multirow{3}{*}{150} & HDPE & \multirow{3}{*}{0.5} & 0.51 & 1.02 \\
\rule[-1ex]{0pt}{3.5ex}  & & Nylon &  & 0.63 & 1.27 \\
\rule[-1ex]{0pt}{3.5ex}  & & Alumina &  & 0.44 & 0.88 \\
\hline
\rule[-1ex]{0pt}{3.5ex}  \multirow{3}{*}{BICEP3} & \multirow{3}{*}{95} & HDPE & \multirow{3}{*}{0.79} & 0.62 & 0.78 \\
\rule[-1ex]{0pt}{3.5ex}  & & Nylon &  & 0.78 & 0.97 \\
\rule[-1ex]{0pt}{3.5ex}  & & Alumina &  & 0.79 & 1.00 \\
\hline
\rule[-1ex]{0pt}{3.5ex}  \multirow{3}{*}{BA4} & \multirow{3}{*}{250} &  HDPE & \multirow{3}{*}{0.3} & 0.58 & 1.93 \\
\rule[-1ex]{0pt}{3.5ex}  & & Nylon &  & 0.72 & 2.41 \\
\rule[-1ex]{0pt}{3.5ex}  & & Alumina &  & 0.39 & 1.28 \\
\hline
\rule[-1ex]{0pt}{3.5ex}  \multirow{3}{*}{BA1} & \multirow{3}{*}{35} & HDPE & \multirow{3}{*}{2.14} & 1.99 & 0.93 \\
\rule[-1ex]{0pt}{3.5ex} & & Nylon &  & 2.15 & 1.01 \\
\rule[-1ex]{0pt}{3.5ex} & & Alumina &  & 4.32 & 2.02 \\
\hline
\rule[-1ex]{0pt}{3.5ex} wSMA & 285 & HDPE & 0.263 & 0.42 & 1.59 \\
\hline
\rule[-1ex]{0pt}{3.5ex} \multirow{2}{*}{CMB-S4 MF} & \multirow{2}{*}{125} & HDPE & \multirow{2}{*}{0.6} & 1.15 & 1.91 \\
\rule[-1ex]{0pt}{3.5ex} & & Nylon &  & 1.3 & 2.17 \\
\hline
\end{tabular}
\end{center}
\end{table}

\section{HIGH FREQUENCY AR COATS} \label{sec:high_freq}
As shown in Table~\ref{tab:AR_params}, the high frequency receivers (BA4 and wSMA) have the thinnest quarter wavelength. This presents a tougher challenge to find materials that can have consistent thicknesses and densities (indexes of refraction) over the area of an optic. The area is particularly challenging for BA4 (200--300 GHz), which has an clear aperture diameter of over 70 cm \cite{Hui2018BA}. The wSMA has a much larger fractional bandwidth, so though the aperture size is much smaller, multi-layer AR coats are required to stay below a reasonable reflection \cite{Grimes2020a}.

Single layer AR coats could work for BA4 but would require strict tolerances on both the index and the density to keep the single null centered. A reasonable tolerance could be plus or minus 5\% off the ideal single layer parameters. For HDPE with an ideal single layer ePTFE coat, the thickness would be 240 $\pm$ 10 $\mu$m and the density $\rho$=1.320 $\pm$ 0.066 g/cc (equivalent to n=1.245 $\pm$ 0.013). Those 5\% tolerances can change the band average reflection by 25\%. All relevant parameters (thicknesses, densities/indexes) are difficult to achieve with traditional high frequency AR materials to the tolerances desired. 

Previous high frequency receivers in the BICEP/\emph{Keck} collaboration have used sintered PTFE (sPTFE) from Porex: the widths of Porex rolls are maximum 33 cm (13 in). Utilizing Porex would require at least two seams to fully cover a BA aperture, which is difficult to work over curved surfaces and prevent potential polarizing effects. And while it is possible to source sPTFE at wider widths than Porex provides, the thicknesses required for high frequencies are not practical to skive to the tolerances required. 

Multi-layer AR coats are more resistant to parameter tolerances and can generate lower reflections than single layer AR coats. For these reasons, we decided to attempt multi-layer AR coats for BA4.

\subsection{DeWAL: Stackable ePTFE}
Expanded PTFE (ePTFE) has been used as AR coats on BICEP3, BA1 and BA2 \cite{bicep3,Dierickx2021}. ePTFE is stretched biaxially from a resin: this produces very small nodules of highly crystalline material surrounded by fibrils of amorphous polymer strands \cite{EbnesajjadSina2016EPAH}. We have shown previously that ePTFE properties can be controlled by compressing it above the PTFE glass transition temperature \cite{Dierickx2021}. 

DeWAL ePTFE membranes (hereafter referred to as DeWAL) can be produced with densities ``ranging from 0.2 to over 1.0 g/cc'' (equivalent indexes of 1.035 to 1.18), thicknesses ``from 0.001 to 0.010 in'' (25.4 to 254 $\mu$m), and widths up to 30 in (76.2 cm) \cite{dewaldatasheet}. These thicknesses and densities are unusually low, and we are unaware of a competitor that produces membranes with a similar combination of properties. The wider widths of DeWAL seem to be limited to the lower density and thinner material, but crucially this ePTFE can be stacked. 

% For some reason, when DeWAL is compressed at temperatures above 50$^{\circ}$C layers will bond. This bonding process works less well at pressures below 3 atm ($\sim$45 psi). If the membrane has already been processed (run through a pressurized temperature cycle, with a bulk PTFE release layer), subsequent attempts at bonding also work significantly less well. Shown in Table~\ref{tab:peel_strength} are 90$^{\circ}$ peel test strengths. We call DeWAL that has never been compressed `Naive' and DeWAL that has been run through a compression cycle at least once `Pre-compressed'. We also report testing of adhesion layer peel strengths: both hot melt adhesives tested have been used as adhesives on cold optics in BA in the past. Low density polyethylene (LDPE) is typically used as the adhesive for other PE optics, such as the vacuum window and the lenses. E100 was used as the adhesive for the BA2 alumina filter AR coat.

When DeWAL is compressed at temperatures above 50$^{\circ}$C layers will bond. This bonding process works less well at pressures below 3 atm ($\sim$45 psi). If the membrane has already been processed (run through a pressurized temperature cycle, with a bulk PTFE release layer), subsequent attempts at bonding also work significantly less well. Shown in Table~\ref{tab:peel_strength} are 90$^{\circ}$ peel test strengths. We call DeWAL that has never been compressed `Naive' and DeWAL that has been run through a compression cycle at least once `Pre-compressed'. We also report testing of adhesion layer peel strengths: both hot melt adhesives tested have been used as adhesives on cold optics in BA in the past. Low density polyethylene (LDPE) is typically used as the adhesive for other PE optics, such as the vacuum window and the lenses. E100 was used as the adhesive for the BA2 alumina filter AR coat.
\begin{table}
    \centering
    \caption{Peel strengths of various kinds of DeWAL compressed together. }
    \begin{tabular}{|M{12cm}|M{3.5cm}|}
\hline
\rule[-1ex]{0pt}{3.5ex} \textbf{DeWAL Layer Recipe} & \textbf{Peel Strength [g/cm]}  \\
\hline
       \rule[-1ex]{0pt}{3.5ex} 2-layer of Naive compressed at 1atm 135$^{\circ}$C & 1.96--3.05 \\
\hline
       \rule[-1ex]{0pt}{3.5ex} 2-layer of Naive compressed at 3atm 135$^{\circ}$C & 3.9--5.9 \\
\hline
       \rule[-1ex]{0pt}{3.5ex} 2-layer of Naive compressed at 10atm 135$^{\circ}$C  & 5.9--7.9 \\
\hline
       \rule[-1ex]{0pt}{3.5ex} 2-layer of Pre-compressed (3 atm) at 1 atm 135$^{\circ}$C & 0.25--0.5 \\
\hline
       \rule[-1ex]{0pt}{3.5ex} 2-layer of Pre-compressed (3 atm) at 1atm 135$^{\circ}$C compressed together with a Naive layer of DeWAL between & 0.15--0.61 \\
\hline
       \rule[-1ex]{0pt}{3.5ex} 2-layer of Pre-compressed (3 atm) at 1atm 135$^{\circ}$C compressed together with 1mil LDPE between & 60--80 \\
\hline
       \rule[-1ex]{0pt}{3.5ex} 2-layer of Pre-compressed (3 atm) at 1atm 135$^{\circ}$C compressed together with 1mil E100 between & 30--40 \\
\hline
    \end{tabular}
    \label{tab:peel_strength}
\end{table}

\begin{figure}
    \centering
    \includegraphics[width=0.5\linewidth]{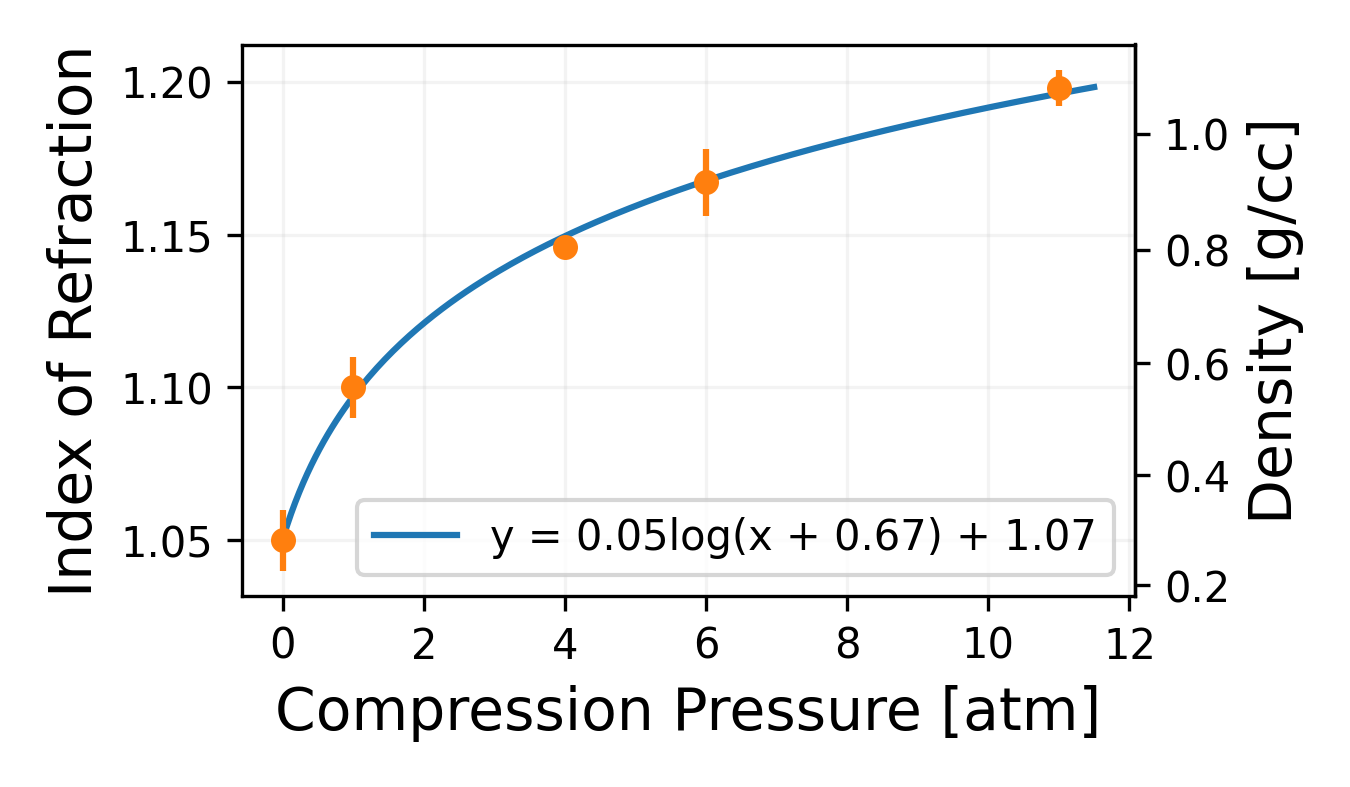}
    \caption{DeWAL density and estimated index at different compression pressures. We estimate the index from the density using the Lorentz--Lorenz relation, assuming that the bulk index and density of PTFE is 1.43 and 2.2 g/cc, respectively. \cite{Polder1946,Lamb1996}}
    \label{fig:DeWAL_compress}
\end{figure}

Similar to other ePTFE AR membranes that we have used, DeWAL changes thickness (and therefore density) when compressed at pressure \cite{Dierickx2021,eiben2022laminate}. Its lower initial density, however, means that even compression at 10 atm does not get the density very high. For example, HDPE's index is around $n_c = 1.55$, so the ideal single layer index is $n_{AR}=\sqrt{1.55}=1.24$, while the highest DeWAL index measured is $\sim$1.2 (see Figure~\ref{fig:DeWAL_compress}). DeWAL therefore requires at least an additional set of layers to push the AR electrical lengths closer to increments of quarter wavelength.

Luckily, with the low index and thickness comes a strength with a multi-layered approach. DeWAL is so low index at the lowest compression that it can act almost like a very thin air gap, which allows for `pseudo-layers' of alternating high and low index materials. Given that typical adhesion materials, such as LDPE, have much higher indexes, the combination of DeWAL and the required adhesive can generate a good AR coat together. To accomplish this we must explore a large parameter space to find the right combination of layer parameters that we believe will produce the most robust AR coat. The next sections explore how we used DeWAL to design AR coats for two high frequency millimeter receivers.

\subsection{Using DeWAL}
\subsubsection{BA4 (200-300 GHz)} \label{sec:BA4}
BICEP Array receiver 4 (BA4) is the highest frequency receiver in the BICEP Array. The frequency band covers from 200 to 300 GHz, which also makes it one of the broader fractional bandwidths covered in Figure~\ref{fig:cheb_comp}. The receiver has three main optical materials that require AR coats: polyethylene (PE) lenses and a window, a nylon filter, and an alumina filter. We will focus on the AR coats for the PE optics, as the nylon filter is a very similar process, while the alumina filter is completely different.

\begin{figure}
    \centering
    \subfloat[]{\includegraphics[width=8cm]{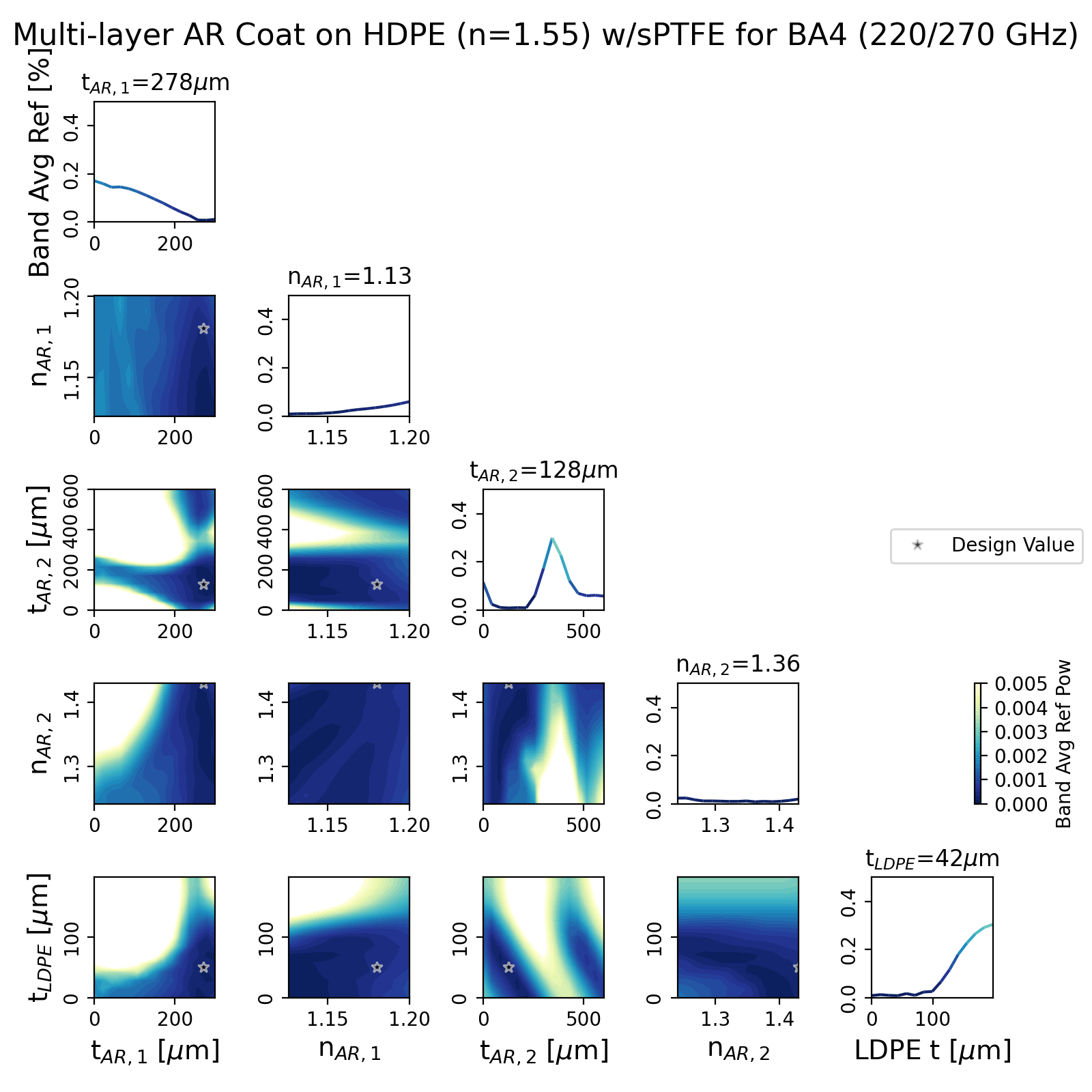} \label{fig:lens_des_BA4}}
    \subfloat[]{\includegraphics[width=8cm]{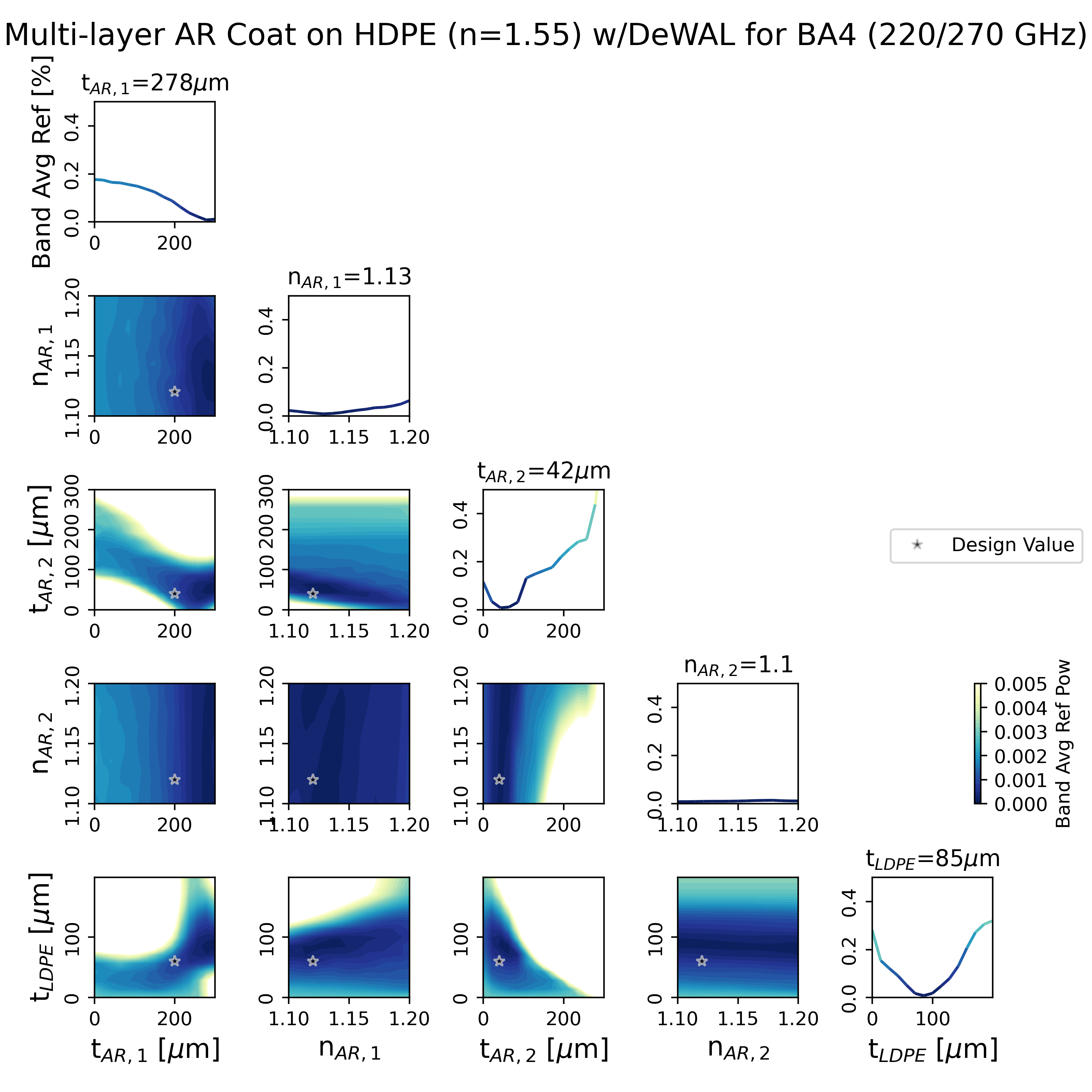} \label{fig:win_des_BA4}}
    \caption{(a) Design corner plot for multilayer High Frequency (HF) HDPE optics using DeWAL, sPTFE, and LDPE layers (BA4, 200--300 GHz). Layers go: AR layer 1 (t$_{AR,1}$,n$_{AR,1}$) stacked DeWAL, LDPE adhesion layer (t$_{LDPE}$), AR layer 2 (t$_{AR,1}$,n$_{AR,1}$) is sintered PTFE, LDPE adhesion layer (t$_{LDPE}$), core. This design was used for the BA4 lenses; stars denote the chosen design parameters. (b) Design corner plot for multilayer High Frequency (HF) HDPE optics using only DeWAL and LDPE layers (BA4, 200--300 GHz). Layers go: AR layer 1 (t$_{AR,1}$,n$_{AR,1}$) stacked DeWAL, LDPE adhesion layer (t$_{LDPE}$), AR layer 2 (t$_{AR,1}$,n$_{AR,1}$) stacked DeWAL, LDPE adhesion layer (t$_{LDPE}$), core. This design was used for the Short Keck window; stars denote the chosen design parameters. Parameters not shown are optimized to produce the lowest possible band average reflection. }
    \label{fig:design_BA4}
\end{figure}

We assume that the index of high density polyethylene (HDPE) is on the higher end of previous measurements at around 1.55 (typical measurements put HDPE indexes between 1.53 to 1.55) \cite{Lamb1996}. This increases the ideal indexes for AR layers slightly, which makes it slightly more difficult to use DeWAL (a low index material). The best solutions are typically pretty similar for the small range of likely HDPE indexes, and therefore if the true index of HDPE is slightly lower than we assume, the AR coat should still work well.

The initial full roll of DeWAL we acquired had a central `mound' of thicker area across the width (see Figure~\ref{fig:dewal_thickness_var} and discussed further in Section~\ref{sec:thick_con}): this limited the minimum thickness that we were willing to use stacked DeWAL. If we used too few layers then we would not be able to rotate the thickness profile enough to average out the problem. Therefore, we resolved to using an initial thick layer of stacked DeWAL (AR1) and some other material for the second layer (AR2). 

From the parameter variation plots (as in Figure~\ref{fig:lens_des_BA4}), there was a clear portion of the high index parameter space that would work for $n_{AR,2}$, all the way up to the index of bulk PTFE (n$\approx$1.43) \cite{Lamb1996}. Skived sheets of bulk PTFE are readily available, if potentially difficult to laminate. Subsequently, we scuffed up the surfaces of the PTFE sheets prior to lamination as much as possible with scour pads. We were unaware, at the time, of an internal review within our group that found that coarse sandpaper works best for roughening bulk PTFE, which we will use in the future. 

\begin{figure}
    \centering
    \subfloat[]{\includegraphics[width=0.7\linewidth]{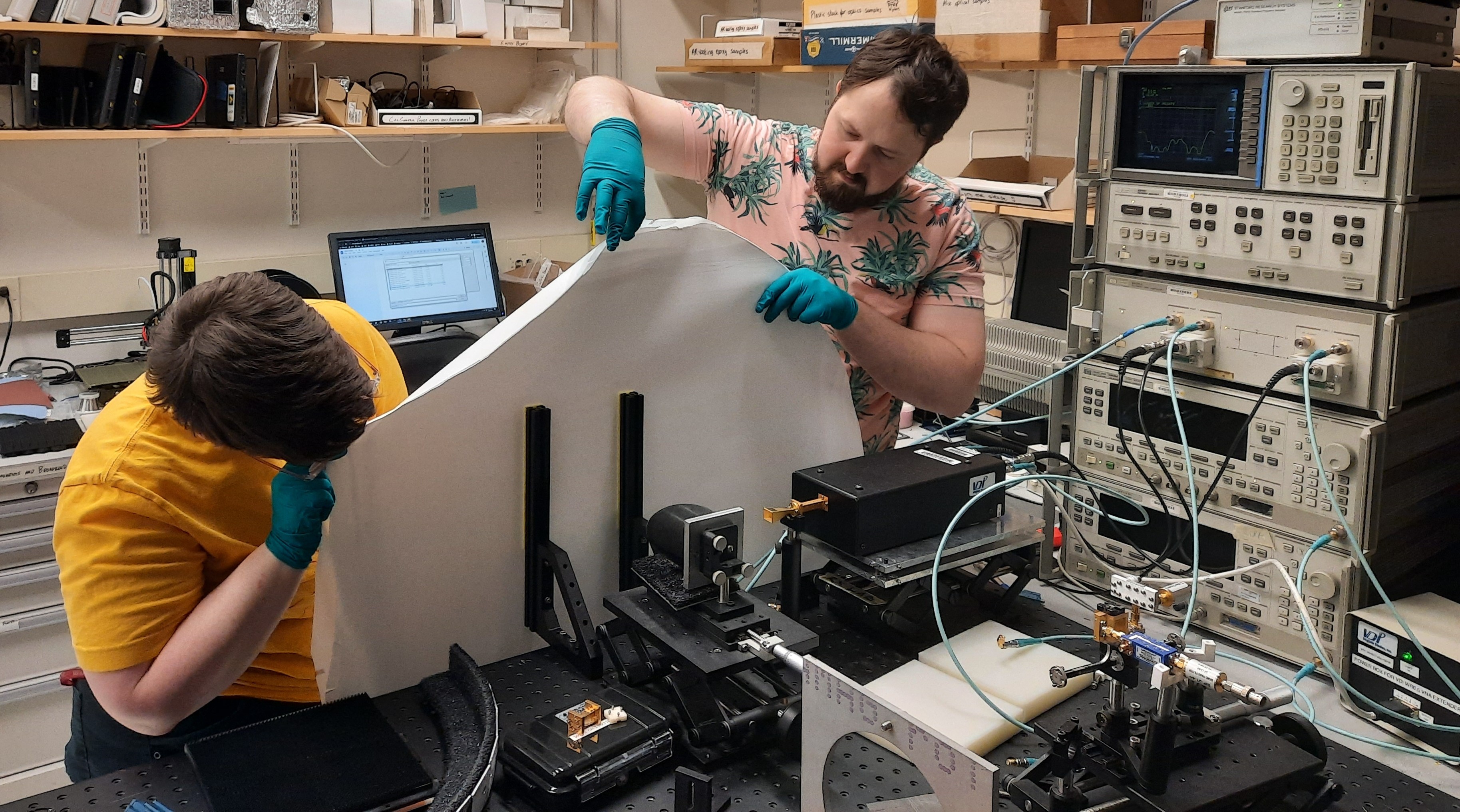}\label{fig:lens_meas_pic}}\par
    \subfloat[]{\includegraphics[width=0.4\linewidth]{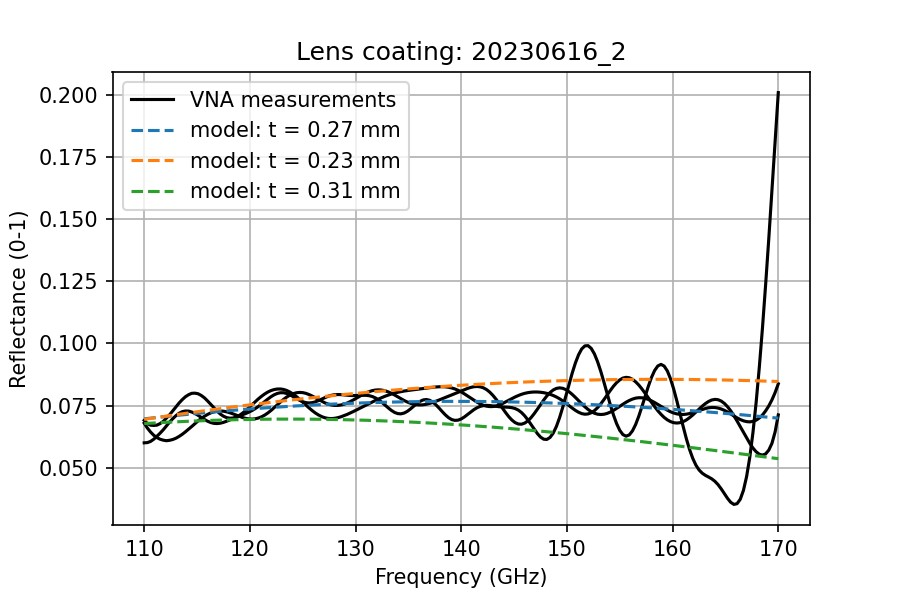}\label{fig:lens_meas_outbnd}}
    \subfloat[]{\includegraphics[width=0.4\linewidth]{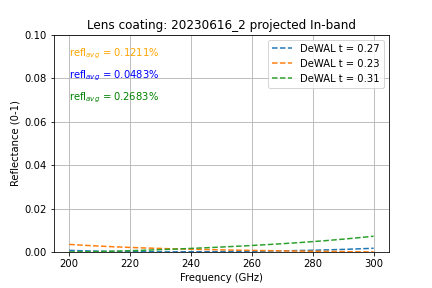}\label{fig:lens_meas_inbnd}}
    \caption{(a) Picture of the out of band measurement of the lens AR coat. The AR coat is held up to the measurement surface by two people (left Miranda Eiben, right Brodi Elwood). Signal from VNA (right most stack of electronics) is sent to a WR6.5 head (center right black box) coupled to free space with a rectangular horn (gold, attached to head). The signal is collimated by a right angle parabolic mirror, the back of which can be seen to the left of the horn. Full scale optics samples are held at the referenced measurement plane with two extruded aluminum posts (center black posts). Photo taken by Matthew Miller. (b) Out of band measurements of a BA4 lens AR coat taken with the VNA with three possible DeWAL (AR1) layer thicknesses. The three modeled thicknesses for the DeWAL layers are 0.23mm, 0.27mm, and 0.31mm (dashed lines). The other layer parameters (such as the thicknesses of the bulk PTFE layer or laminating LDPE) are assumed to match the physical measurements taken pre-lamination. (c) The same parameters in the models from (b) are modeled on an infinitely thick  core of HDPE to judge efficacy of AR parameters.}
    \label{fig:lens_meas}
\end{figure}

We optically tested the AR coating for the lenses before laminating them to the lens surface. These measurements were taken with a Vector Network Analyzer with a Virginia Diodes WR6.5 head. The signal is coupled to free space with a WR6 rectangular gain horn. The beam is then collimated with a 90$^{\circ}$ off axis parabolic mirror, with the beam waist placed where two posts allow us to hold large optical materials up to the flat surface the posts provide. We normalize the measurement with a short measurement of a flat aluminum plate at the measurement surface (see Figure~\ref{fig:lens_meas_pic}). Time gating the S11 (reflection) signal (calculated by our HP 8510C Network Analyzer) allows us to remove spurious reflections from the system, such as those from the horn. 

Figure~\ref{fig:lens_meas_pic} is a picture of a measurement with this set up in progress of the AR coating for a lens. Figure~\ref{fig:lens_meas_outbnd} shows the actual VNA measurements of the lens AR coat; at this point, all layers are adhered together, except for the final LDPE adhesion layer to the lens. Therefore, the stack is designed to be: DeWAL (t$_{AR,1}$=0.27, n$_{AR,1}$=1.18), LDPE adhesion (t$_{LDPE}$=0.05 (2 mil), n$_{LDPE}$=1.51), and bulk PTFE (t$_{AR,2}$=0.127 (5 mil), n$_{AR,2}$=1.43), as shown in Figure~\ref{fig:lens_des_BA4}. We also plotted models with three potential thicknesses of the DeWAL layer (dashed lines), to test the tolerances of our stack. In Figure~\ref{fig:lens_meas_inbnd} we project the modeled AR layers, with an infinitely thick HDPE core, into the design band. Also, we take averages of the reflected power in band and find that this multi-layer AR coat is expected to produce band average reflections between 0.05\% and 0.27\%; this is consistent with the predictions from the design corner plots in Figure~\ref{fig:lens_des_BA4}. 

% \begin{figure}
%     \centering
%     \includegraphics[width=8cm]{lens_ar_measurement.jpg}
%     \subfloat[b]{\includegraphics[width=6cm]{lens_20230616_2_vnarefplot.png}}
%     \subfloat[c]{\includegraphics[width=6cm]{lens_20230616_2_refplot.png}}
%     \caption{Caption}
%     \label{fig:enter-label}
% \end{figure}

The initial thin window for BA4 used a very similar design of AR coating. In retrospect this was a mistake: the bulk PTFE was too stiff to deform with the window, and during initial pump down the AR coating delaminated. We had to return to the drawing board. At this point, we had aquired a new roll of DeWAL, which had much better thickness consistency across the width (see Figure~\ref{fig:dewal_t_var_new}). This allowed us to explore using only DeWAL for a multi-layer AR coat, as thinner layers could be made with more consistent thickness across the entire span.

Therefore we explored a new portion of the parameter space in the second design corner plot in Figure~\ref{fig:win_des_BA4}: now the second AR layer (AR,2) is a DeWAL layer, so the parameters are bounded by the physical parameter constraints set by DeWAL. The thickness is set to maximum 300 $\mu$m, to limit the number of layers required, and the indexes are limited to between 1.1 and 1.2, as it is most practical to compress DeWAL to those densities. The design chosen from these parameters (as shown by the stars in Figure~\ref{fig:win_des_BA4}) was tested on a smaller bulk HDPE vacuum window for a test cryostat called Short Keck.

We first compressed the outer layer (AR,1), as that was the thickest layer of DeWAL required. The stack was designed to be 10 layers of DeWAL compressed at 3 atm (45 psi), attempting to get the layers to be approximately 280 $\mu$m thick. The resultant stack was approximately 200 $\mu$m thick (80 $\mu$m too thin), and as shown in Table~\ref{tab:peel_strength} adding additional layers to a DeWAL stack does not work. We had to redesign the rest of the layers to compensate for the missed specification of the outer layer.

Thankfully, the design corner plots can help point to a set of parameters for the other layers that will work with a different parameter for a layer. Following the first column of Figure~\ref{fig:win_des_BA4} (t$_{AR,1}$), we can see that decent multi-layer AR coat lies with a slightly thinner layer of LDPE (from 85 $\mu$m to about 50 $\mu$m) and a thicker mid AR layer (from 42 $\mu$m to 90 $\mu$m). We can rework the design of the other layers during the processing to account for these different layers. 

\begin{table}[]
    \centering
    \caption{Parameters used for the BA4 band optics (BA4 lenses in Figure~\ref{fig:lens_meas_inbnd}, SK window in Figure~\ref{fig:sk_wSMA_meas_ref}). The layers go air, AR1, LDPE, AR2, LDPE, Core. }
    \begin{tabular}{|P{4.75cm}|P{2.75cm}|P{1.5cm}|}
         \hline
\rule[-1ex]{0pt}{3.5ex} \textbf{BA4 Lens AR Layers} & \textbf{Thickness [$\mu$m]}& \textbf{Index} \\
         \hline
\rule[-1ex]{0pt}{3.5ex} AR,1 & 230, 270, 310 & 1.18 \\
         \hline
\rule[-1ex]{0pt}{3.5ex} LDPE & 50 & 1.51 \\
         \hline
\rule[-1ex]{0pt}{3.5ex} AR,2 & 127 & 1.43 \\
         \hline
\rule[-1ex]{0pt}{3.5ex} \textbf{SK Window AR Layers} & \textbf{Thickness [$\mu$m]}& \textbf{Index} \\
         \hline
\rule[-1ex]{0pt}{3.5ex} AR,1 & 200 & 1.12 \\
         \hline
\rule[-1ex]{0pt}{3.5ex} LDPE & 60 & 1.51 \\
         \hline
\rule[-1ex]{0pt}{3.5ex} AR,2 & 40 & 1.12 \\
         \hline
    \end{tabular}
    \label{tab:Sk_parameters}
\end{table}

The final design of the Short Keck (SK) slab window AR coat was meant to act as a test for the full scale BA4 thin window. Short Keck is a test cryostat for BICEP Array detector validation: a vacuum window with appropriate AR coat reduces confounding optical inefficiencies within the cryostat. Therefore, SK also requires the same bandpass as BICEP Array.  The SK slab window was sufficiently thick (approximately 38 mm) that the time gating in the previously described WR6.5 free space VNA (see Figure \ref{fig:lens_meas_pic}) allowed us to only observe the reflections off a single surface. These out of band (110--170 GHz) measurements are shown with a fit that predicts what the reflection will be in band.

% \begin{figure}
%     \centering
%     \includegraphics[width=0.5\linewidth]{220_270ghz_hdpe_doubleSK_Mod_AR.png}
%     \caption{Caption}
%     \label{fig:SK_tol}
% \end{figure}

To make a better window AR coat, we need to be either able to hit a thicker target for the outer DeWAL layer, or increase the thickness of the second AR layer. This should bring the top edge of the reflections in band down, further reducing the band average reflection. 

\begin{figure}
    \centering
    \includegraphics{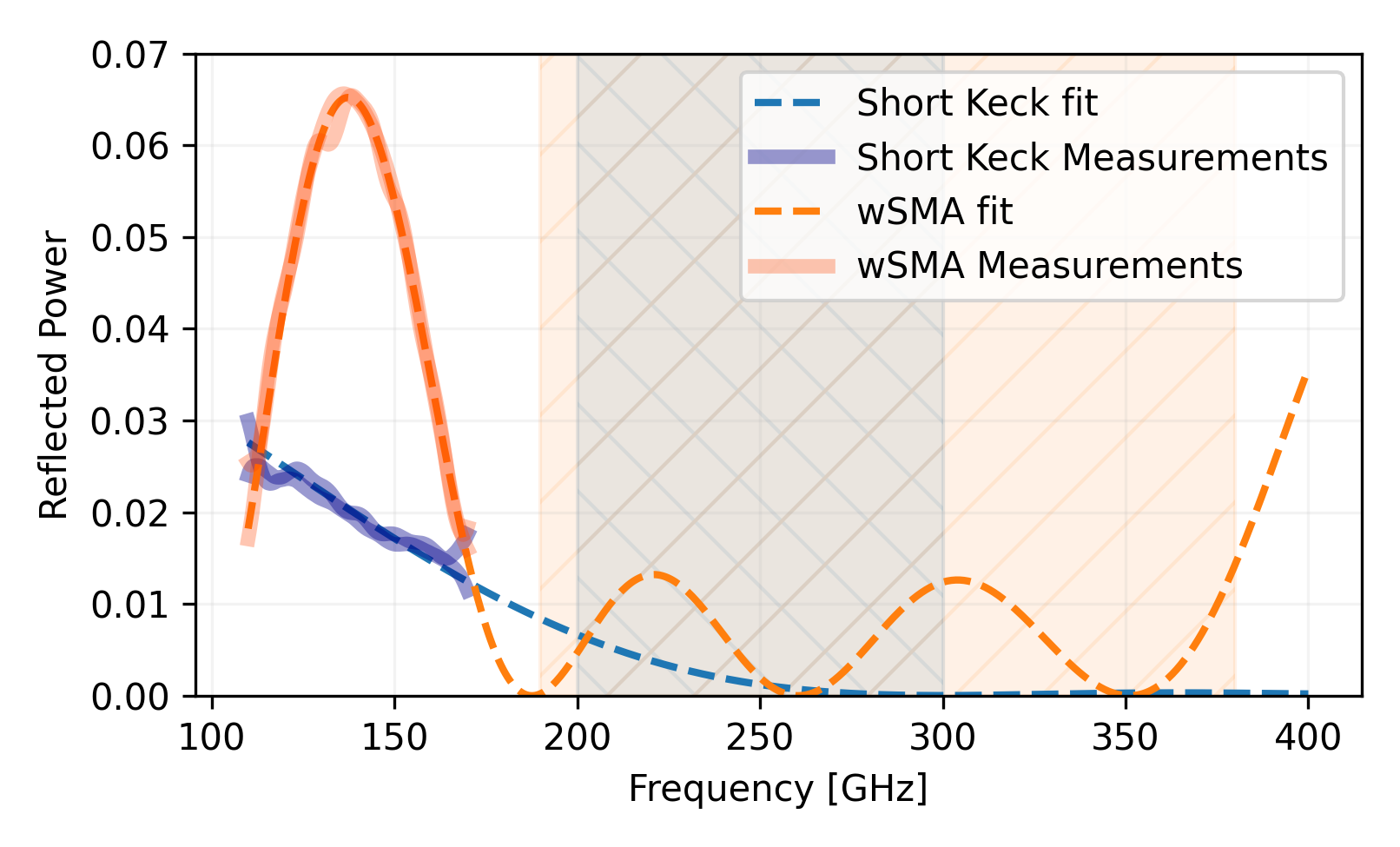}
    \caption{Modeled and measured anti-reflection coats for a slab window for high frequencies (Short Keck) in blue and a thin window (wSMA) in orange. The bands are similarly color coded. Solid lines are measurements, dashed lines are fits. The slab Short Keck measurements are a single incidence reflection, while the wSMA measurements are reflections off all layers. The projected band average reflected powers are modeled to be 0.6\% for the wSMA window, and 0.2\% for the Short Keck window.}
    \label{fig:sk_wSMA_meas_ref}
\end{figure}

\subsubsection{wSMA (190--380 GHz)} \label{sec:wSMA}
The Sub-Millimeter Array (SMA) is a interferometer on Mauna Kea. The receivers in the array are currently being upgraded to a wider bandwidth (wSMA) \cite{Grimes2020a}. The new receivers have a handful of transmissive optics, such as dichroic plates, an IR filter, and a vacuum window. Naturally, as the bandwidth is expanding, the AR coats for the transmissive optics must be redesigned. For this proceedings, we will focus on the thin high modulus polyethylene (HMPE) vacuum window.

The wSMA window was designed after the BA4/SK window described in Section~\ref{sec:BA4}. As such, the window was designed with using the same materials as were put on the SK window, as there would be plenty of offcuts. We explored the adhesion layer parameter space, with set DeWAL thicknesses. As the deeper layers of the coat prefer thinner amounts of DeWAL, we tested thickness increments of DeWAL stack, with a preference for combinations of layers that did not require additional precompression. The final design utilizes a single layer of DeWAL (t = 20 $\mu$m) as a mid DeWAL layer, the only new layer of DeWAL needed for this design.

\begin{figure}
    \centering
    \includegraphics[width=10cm]{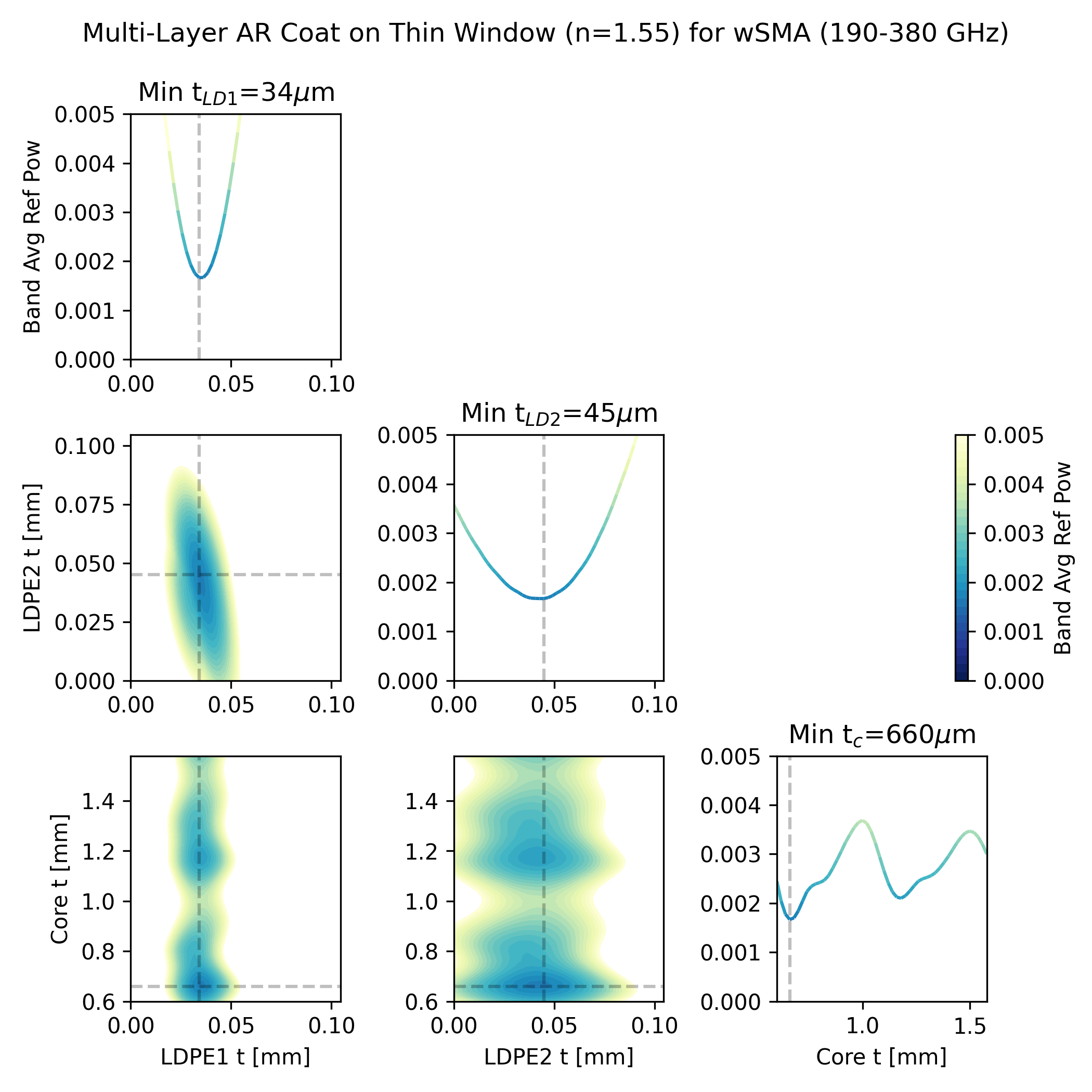}
    \caption{Design corner plot for a wSMA thin window with three DeWAL layers, two LDPE layers. The DeWAL layers had set thickness ($t_{DW1}=200 \mu$m, $t_{DW2}=20 \mu$m and $t_{DW3}=40 \mu$m) and indexes of 1.13. The LDPE has an assumed index of 1.51.}
    \label{fig:wSMA_design}
\end{figure}

The preferred thicknesses of LDPE are not necessarily commercially available, which can also influence the design. In this case, the first LDPE layer's best thickness of 34 $\mu$m is close to 1.25 mil (32 $\mu$m) LDPE and the second's best thickness of 45 $\mu$m 2 mil (50 $\mu$m). These minor thickness changes do not affect the band average reflection much, particularly because the changes move roughly along the correlated vector ($t_{LD1}$ goes down, while $t_{LD2}$ goes up). This can be observed in the contour map of the minimum parameter space of the two LDPE layers in Figure~\ref{fig:wSMA_design}. 

The fully AR coated window was measured in the same WR6.5 VNA setup shown in Figure~\ref{fig:lens_meas_pic} and described in Section~\ref{sec:BA4}. The measurements are plotted in Figure~\ref{fig:sk_wSMA_meas_ref}, which show the out of band reflections off all layers (i.e., both mirrored AR layers and the core). The core thickness was a little higher than the design preferred (approximately 700 $\mu$m instead of 660 $\mu$m), which resulted in slightly higher band average reflection. One of the design goals was to keep the reflected power magnitudes below $-20$ dB (approx 1\%), which this window did not accomplish. We likely can produce a better in-band reflection profile by adding more layers of LDPE/DeWAL pseudo-layers, as the current combination of layers produces an electrical length that is not close to an even multiple of quarter wavelengths (see Table~\ref{tab:AR_params}).

\subsection{Problems with Using DeWAL}
As one may expect from developing new techniques with a new material, we have encountered some difficulties with using DeWAL for AR coats. We describe a few of the problems we have encountered here, such that the reader may be aware of the potential issues that must be overcome to use this material. These problems do not detract from the usefulness of such a low density and low thickness material for multi-layered AR coats but are meant to advise readers what they may be getting into when using it.

\subsubsection{Bonding process}
As shown in Table~\ref{tab:peel_strength}, the bonding process between layers seems to only work between two Naive layers; when attempting to add additional layers to an already compressed stack, either Pre-compressed or new Naive layers will not bond as well as two layers of Naive DeWAL will bond to each other. Even 1 atm compressed Naive layers exhibit 10 times stronger bonding than Pre-compressed layers with each other.

We are unsure why the bonding seems processing specific: it is possible that the heated compression changes the surface texture or the receptiveness of the polymer chains to making weak bonds with other chains. However, practically this means that a stacked layer of DeWAL membranes must be made in one compression cycle. If additional electrical length is required to meet reflection requirements, they must be added with hot melt adhesive layers. 

\begin{figure}
    \centering
    \subfloat[]{\includegraphics[height=6cm]{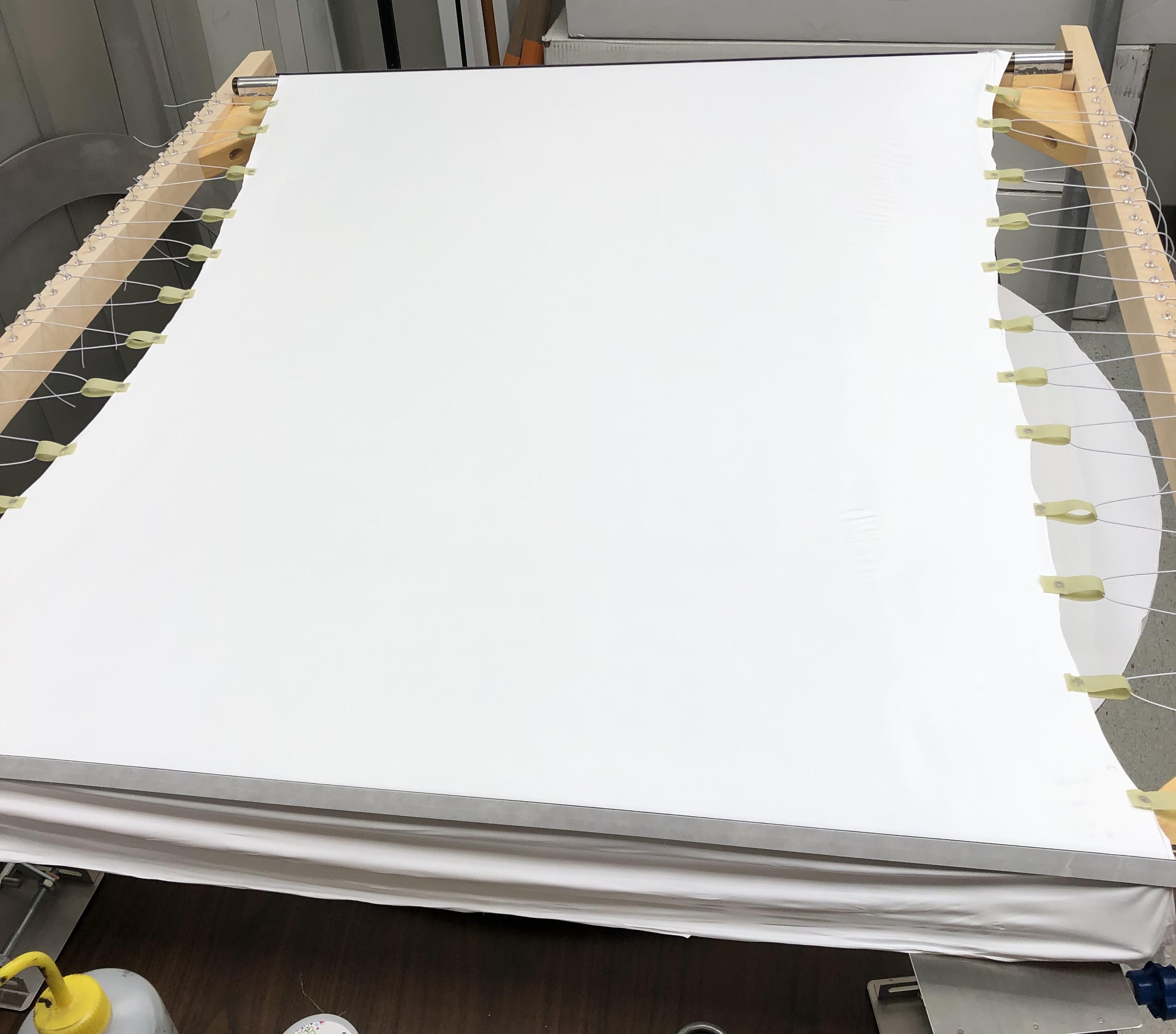}\label{fig:stretch_rig}}
    \subfloat[]{\includegraphics[height=6cm]{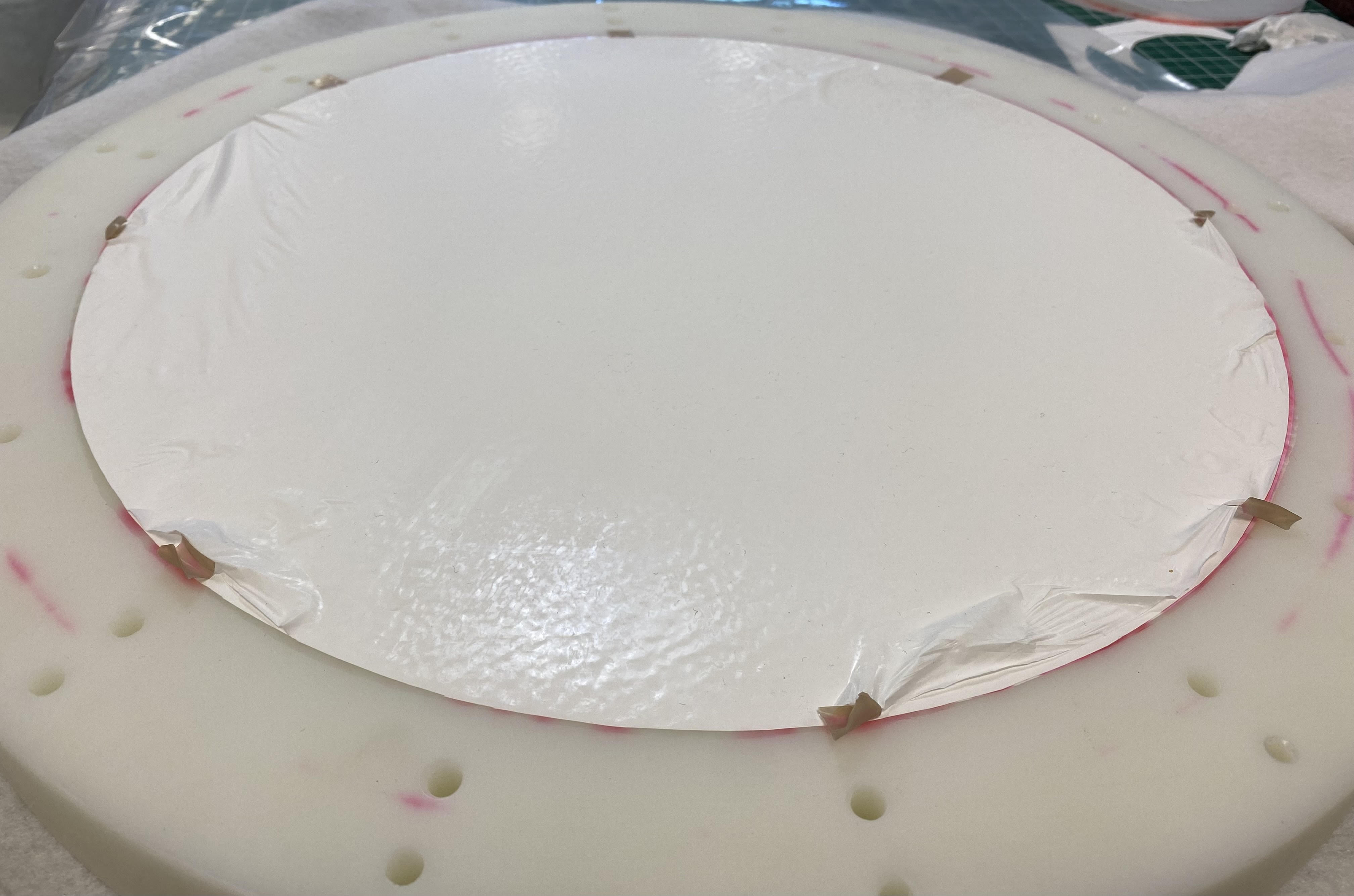}\label{fig:peeled_dewal}}
    \caption{(a) DeWAL stretching rig, which flattens a sheet of DeWAL before it is smoothed over a flat plate. This process helps prevent wrinkles and bubbles from forming under a sheet, which must be smoothed out by hand. (b) The AR coating of the Short Keck (SK) window, with the top layer of DeWAL peeled up. The top layer got caught on the release layer and came up when the release layer was removed.}
    \label{fig:dewal_pics}
\end{figure}

\subsubsection{Static}
DeWAL, as with most ePTFEs, is incredibly prone to accumulating surface static. This generally means that, unless handled in clean room environments, the AR layers will attract dust in large, visible quantities. The static can be mitigated somewhat with ionizing fans, but thankfully small bits of dust and fibers do not seems to adversely effect AR performance. Copious amounts of dust may impact bonding performance, however, so static should be mitigated as much as possible.

However, the static does introduce difficulties in handling cut sheets of DeWAL, both big and small. Because of the membrane's incredibly low density, the static can generate enough force to move the DeWAL. If holding a long sheet up by an end and letting it hang down, it will be attracted towards a person's body. If laying small sheets onto another static prone surface, such as the bulk PTFE release layer, the DeWAL may shift, or lift off the surface. It is also difficult to blow an ionizing fan over uncompressed DeWAL due to the air flow potentially lifting up the material.  

Therefore, handling long sheets are best accomplished with two people or a rig, such that four corners of the sheet can be secured and prevented from moving (see Figure~\ref{fig:stretch_rig} for a picture of our DeWAL stretching rig). Additionally, we have found that 92-600 Nitrile texture-less gloves work best when smoothing bubbles and wrinkles by hand. Other types of gloves tend to catch on the DeWAL, causing it to bunch up and wrinkle. 
\begin{figure}[!t]
    \centering
    \subfloat[]{\includegraphics[width=0.45\linewidth]{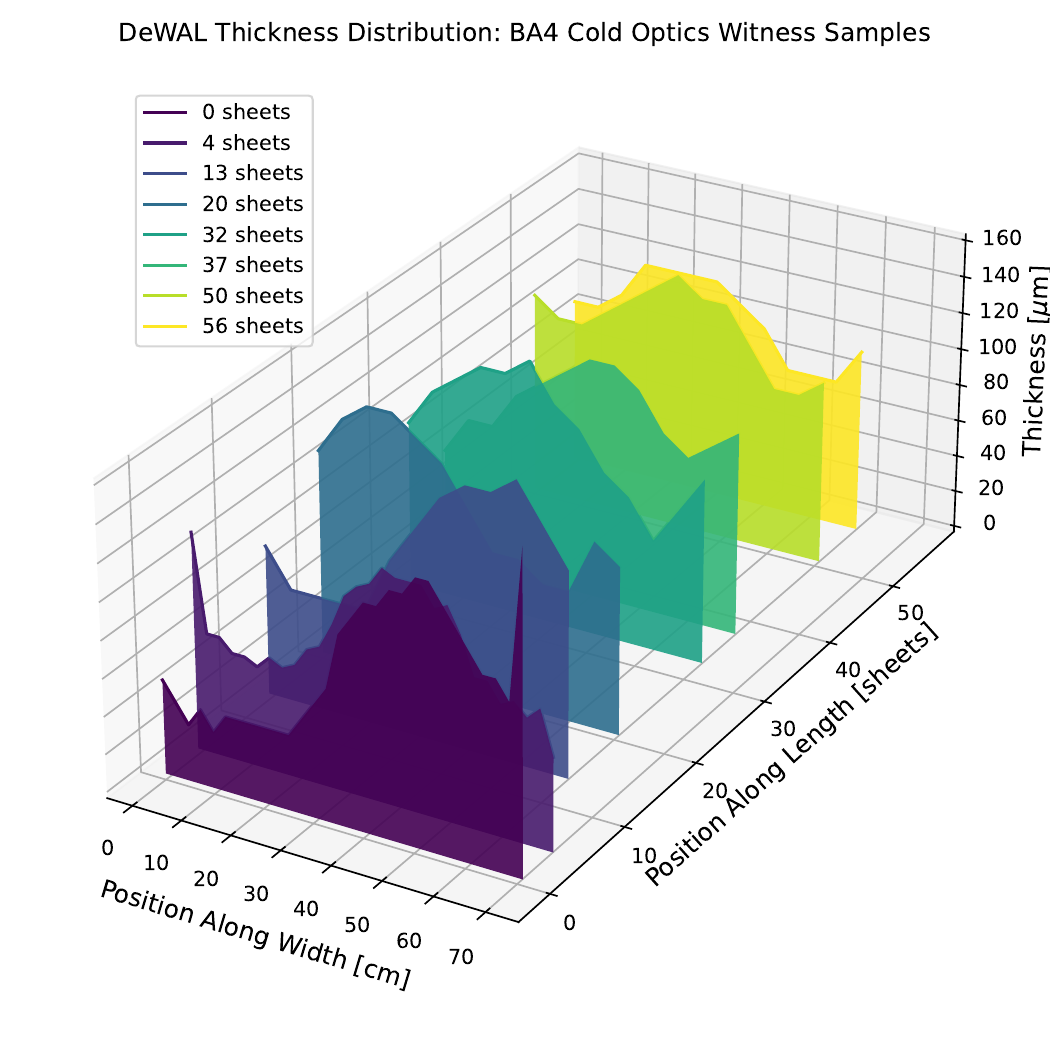}\label{fig:dewal_t_var_init}}
    \subfloat[]{\includegraphics[width=0.45\linewidth]{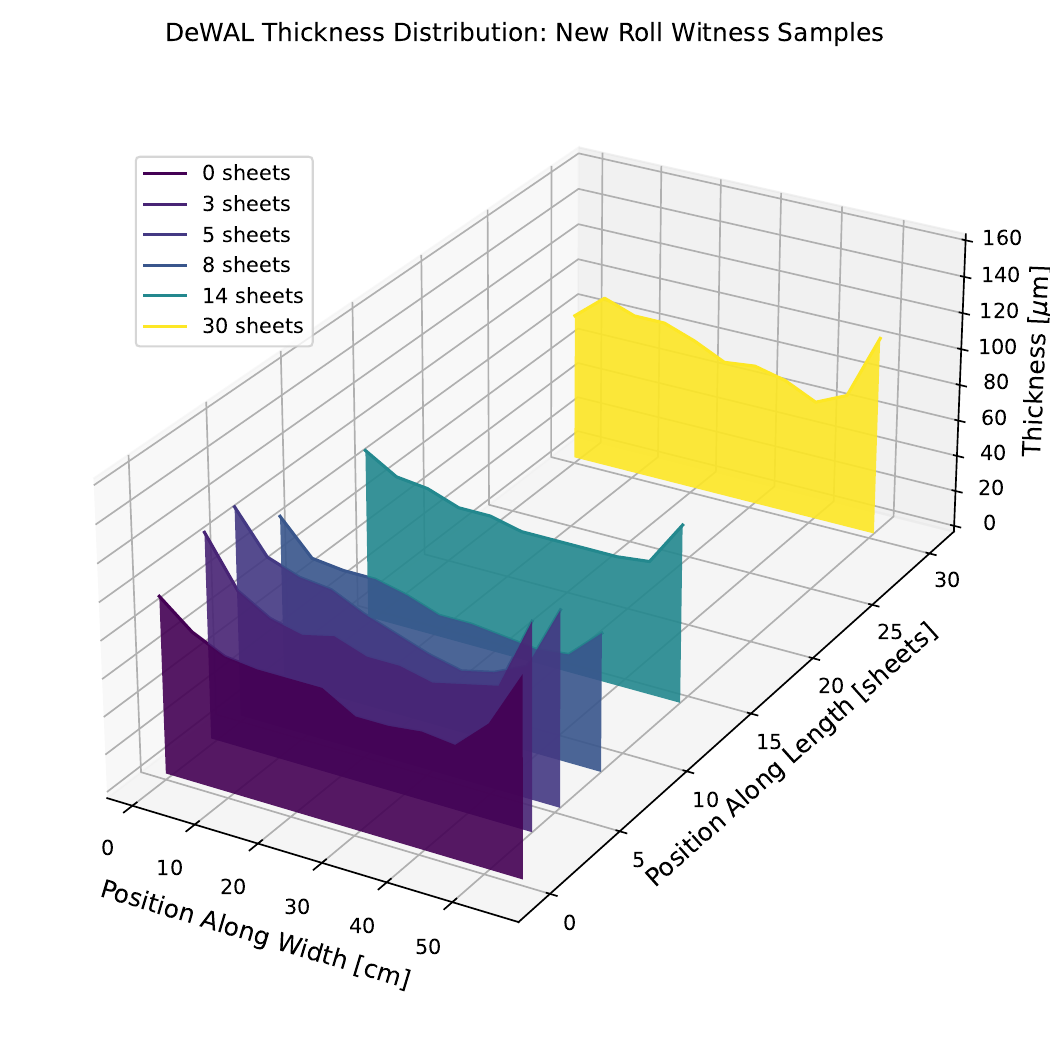}\label{fig:dewal_t_var_new}}
    \caption{(a) Thickness distribution across width over the roll of DeWAL used for the BA4 cold optics. Length along the roll is measured in `sheets', which are roughly 75 cm lengths cut off the roll. (b) Thickness distribution across width over the roll of DeWAL used for the SK window and wSMA window. Length along the roll is measured in `sheets', which are roughly 75 cm lengths cut off the roll.}
    \label{fig:dewal_thickness_var}
\end{figure}

\subsubsection{Release layer}
When compressing the DeWAL, or adhereing it to another surface, there must be a bulk PTFE release layer above the layers. DeWAL tends to stick to other release layers (such as polyester). Even bulk PTFE can generate enough static force to peel up the top layer of DeWAL; release must be done carefully, in small increments, preferably with an ionizing fan running over the peel surface. See Figure~\ref{fig:dewal_pics} for an example of an AR coat that had a DeWAL layer peel off with the release layer. The authors suspect that the layer came up with the release layer because of improper cooling after adhesion to the slab window: the bottom of the slab was still warm, and therefore there was uneven contraction of the release layer, which pulled on the DeWAL layer. 

\subsubsection{Thickness consistency} \label{sec:thick_con}
DeWAL comes in rolls at least 200 feet long. We then cut the appropriately scaled `sheets' off these rolls and lay the layers onto a flat surface with a release layer, before transferring that stack into our optics autoclave \cite{eiben2022laminate}. We typically rotate the stack between layers of DeWAL, to somewhat evenly distribute any thicknesses profile across the stack. However, this was complicated somewhat by the initial roll having a changing thicknesses profile across the roll. 

In Figure~\ref{fig:dewal_t_var_init}, we show how the thickness varied across the initial roll of DeWAL, which was used for the BA4 cold optics. As shown, the `mound' of the thickest section moved back and forth along the width of the roll. This presented a problem with making the final stacked AR coating a consistent thickness. The latest roll had much better thickness consistency across the roll, but was about half as thick initially as that center mound. The thickness distribution across the new roll is shown in Figure~\ref{fig:dewal_t_var_new}, which was used for the Short Keck window and wSMA windows (described in Sections~\ref{sec:BA4} and \ref{sec:wSMA}). 

The production company has been very receptive to attempting new techniques to try to keep better tolerances, so the authors hope that the next rolls will only improve their usefulness.

\section{NEXT STEPS}
\subsection{BA4 Window}
The next science grade BA4 window will be a thin high modulus polyethylene (HMPE) window with a purely DeWAL and LDPE AR coat. It will be similar to the design tested on the slab Short Keck window (design shown in Figure~\ref{fig:win_des_BA4}, measurements shown in Figure~\ref{fig:sk_wSMA_meas_ref}), though with an additional design parameter for the core window thickness. The window thickness is likely to be approximately $\lambda$ or about 1.2 mm (1200 $\mu$m) thick.

This thin window is anticipated to deploy with the BA4 receiver in the austral summer 2024/2025, where it will join previously deployed thin windows on BA2 (150 GHz) and BICEP3 (95 GHz). We will be reporting the results from the first year of observations with thin HMPE windows in an upcoming paper.

\subsection{wSMA Window}
As discussed in Section~\ref{sec:wSMA}, the current design AR coated thin window will perform better with a slightly thinner window. We will be testing that design in the coming months. There other combinations of parameters that may work better than the current design, such as a set of layers that generate an equivalent double layer AR coat. As the current design was meant to use pre-generated stacks, opening the parameter space will likely find a more tolerant set of layers.

\subsection{CMB-S4}
\begin{figure}[!t]
    \centering
    \includegraphics[width=0.5\linewidth]{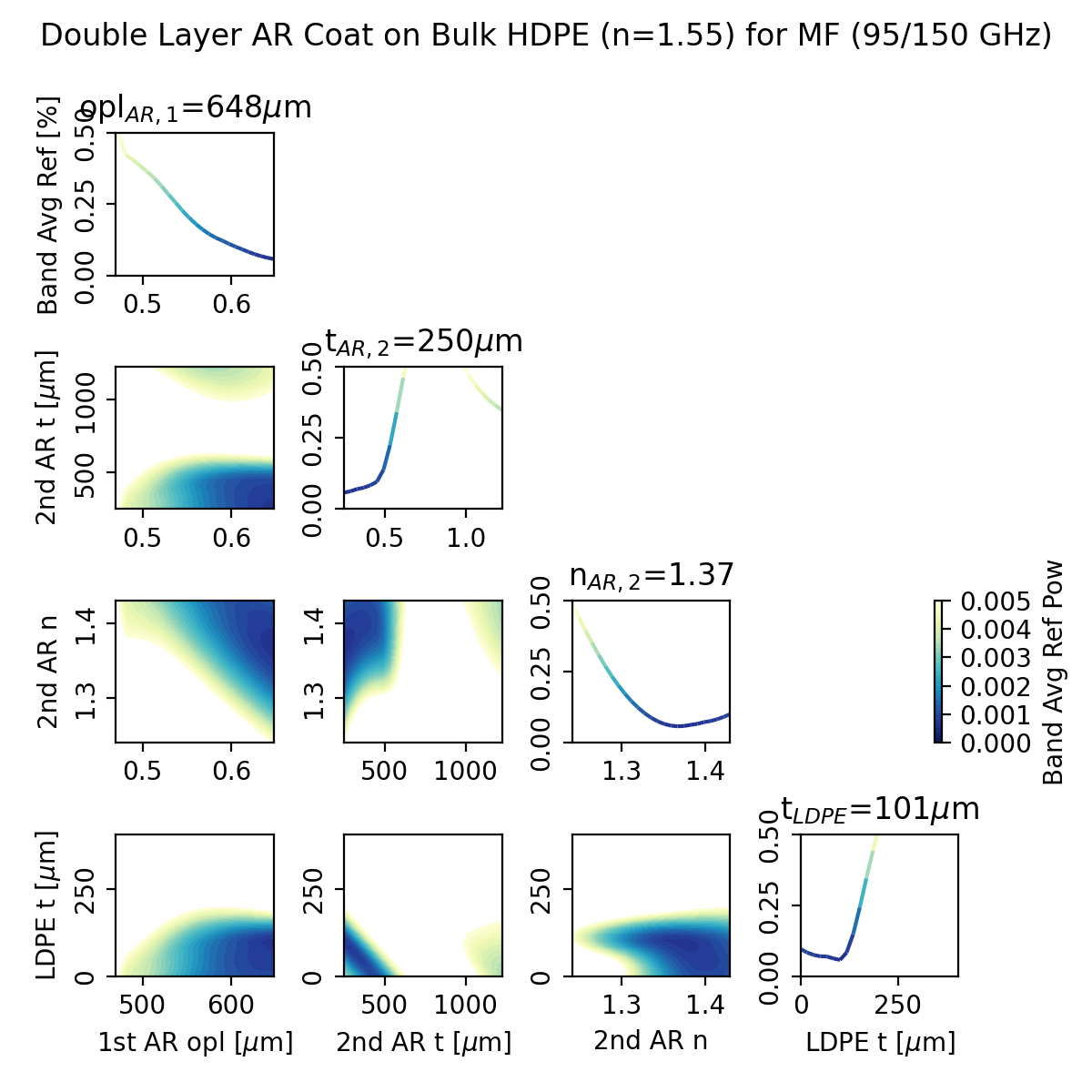}
    \caption{Design corner plot for multilayer Mid Frequency (MR) HDPE optics using SH Teadit and a skived sPTFE. \cite{Dierickx2021} The layers are modeled as the 1st AR optical path length (opl) using SH Teadit, then an LDPE adhesion layer, then the 2nd AR (t and n) being sPTFE, and a final LDPE adhesion layer to the infinitely thick optic. Parameters not shown are optimized to produce the lowest possible reflection.}
    \label{fig:cmb-s4_design}
\end{figure}

CMB Stage 4 (CMB-S4) is the next generation of CMB receivers currently being designed. The CMB-S4 mid-frequencies (MF) are dual band 95/150 GHz and have the widest fractional bandwidth studied in this proceedings. Other CMB-S4 bands are very similar to previously designed BICEP/\emph{Keck} bands; the high frequency (HF) CMB-S4 bands, for example, are very similar to the BA4 band \cite{abazajian2019cmbs4}. 

The proposed layered AR coat design in Figure \ref{fig:cmb-s4_design} for the CMB-S4 Mid-frequencies utilizes two previously tested materials. Compressed ePTFE and a higher density sPTFE skived to the appropriate thickness, can produce a near ideal double layer. As the frequencies are lower, the skived sPTFE thickness is large enough to be producible, unlike the required thickness at high frequencies.  

Though this particular stack of layers has not been tested together, the individual components have been tested separately \cite{Dierickx2021}. We expect that this design is eminently achievable as the thickness tolerance at lower frequencies tend to be more lax.

\subsection{Other Uses for DeWAL}
DeWAL, as a low density low thickness membrane, could be used for other optics. Physically separated sheets of uncompressed DeWAL could potentially be used as a Radio Transparent Multi-layer Insulation (RT-MLI) \cite{Choi_2013}. DeWAL's very small pore size (likely similar to other ePTFEs at around 12 $\mu$m) could eliminate the scattering concerns that are raised by the approximately 400 $\mu$m cell sizes that the current BA RT-MLI material \cite{goldfinger2022thermal}.

A many layered (on order tens of layers) stack of DeWAL and LDPE could be used to generate a low pass edge filter. Developing this filter would likely require an easier procedure to lay DeWAL flat for these many layers such that it is not egregiously time-consuming to generate.

\section{CONCLUSIONS}
Generating broadband anti-reflection coatings is a difficult design problem. With a large number of potential physical parameters, the multi-variate design space is hard to visualize and ensure that the best potential solutions are being found. As such, we propose two different methods for testing design solutions.

We report the theoretical lower limit that multi-layer AR coats can achieve over a fractional bandwidth. These ideal AR coats parameters are found using Chebyshev polynomials to space the destructive `nulls' in frequency space; this produces equal ripple frequency dependent reflection profiles. We compare the achieved band average reflected powers from the latest generation of BICEP/Keck experiments' AR coats, and proposed AR coats for the wSMA and CMB-S4 MF experiments.

Broadband high frequency mm-wavelength AR coats provide the greatest technical challenge for layered coats due to the unusual design constraints for these bands. Finding a coating material with the correct electrical length (combined index and physical thickness) within a reasonable tolerance over the large aperture sizes of new mm receivers is very difficult. Therefore, multi-layered AR coats are required to reduce tolerance constraints. We demonstrate a technique that allows us to fully probe the physically possible parameter space for multi-layer AR coats. By plotting which parts of the multi-variate the parameter space produce the lowest band average reflection, one can see directly which combination of parameters produce a good AR coat.

We characterize a new expanded polytetraflouroethylene (ePTFE) membrane for multi-layer AR coatings called DeWAL. DeWAL exhibits an interesting behavior that allows the thin, low density layers to be stacked together, generating a thicker and higher density sheet. These stacked sheets of DeWAL can be used to generate good high frequency AR coats in combination with other layers such as LDPE and bulk PTFE. DeWAL has now been used successfully for two different high frequency AR coats. The AR coats for the lenses, nylon filter and vacuum window in BA4 (200-300 GHz) all use different designs using compressed DeWAL stacks. The AR coat for the vacuum window for the wSMA (190-380 GHz) also uses multi-layer, DeWAL and LDPE only AR coat. 

These multi-layer AR coatings are not the only potential use cases for a thin, low density membrane, just the only ones explored here. DeWAL has the potential to be used for a variety of filter applications. 

The authors thank the DeWAL team (particularly Al Horn, Tom Palasky, and Ray Patrylak) for their generous gifts of samples, data, experience, and time.

% References
\bibliography{report} % bibliography data in report.bib
\bibliographystyle{spiebib} % makes bibtex use spiebib.bst

\end{document}